\documentclass[conference]{IEEEtran}

\IEEEoverridecommandlockouts

\usepackage{cite}
\usepackage{amsmath,amssymb,amsfonts}
\usepackage{algorithmic}
\usepackage{graphicx}
\usepackage{textcomp}
\usepackage{xcolor}
\usepackage{hyperref}
\usepackage{enumerate}

\usepackage{comment} 
\usepackage{kotex} 

\newcommand{\YOON}[1] {
	\textcolor{red}{\bfseries{YOON: {#1}}}
}

\newcommand{\Skip}[1] {
}

\def\BibTeX{{\rm B\kern-.05em{\sc i\kern-.025em b}\kern-.08em
    T\kern-.1667em\lower.7ex\hbox{E}\kern-.125emX}}

\title{Collaborative 3D modeling system based on blockchain}

\author{\IEEEauthorblockN{Hunmin Park}
\IEEEauthorblockA{\textit{School of Computing} \\
\textit{KAIST}\\
Daejeon, Korea \\
95phm@kaist.ac.kr}
\and
\IEEEauthorblockN{Sung-Eui Yoon}
\IEEEauthorblockA{\textit{School of Computing} \\
\textit{KAIST}\\
Daejeon, Korea \\
sungeui@kaist.edu}
}

\newcommand{\myParagraph}[1]{\noindent\textbf{#1}}

\newcommand{\myEmph}[1]{\emph{#1}}

\begin{document}
\maketitle

\begin{abstract}
We propose a collaborative 3D modeling system, which is based on the blockchain
technology. Our approach uses the blockchain to communicate with modeling
	tools and to provide them a decentralized database of the mesh
	modification history. This approach also provides
a server-less version control system: users can commit their modifications to
	the blockchain and checkout others' modifications from the blockchain.
	As a result, our
system enables users to do collaborative modeling without any central server.
\end{abstract}


\section{Introduction}

With the development of the game and animation industries to meet the desire
and expectation of consumers, we have to deal with large and complex 3D models.
To build large and complex models, collaboration among many designers becomes the common practice.
Unfortunately, modeling tools like Blender does not support collaborative
modeling yet due to its complexity. Instead, we can use a separate, version
control system (VCS) such as Git, but there is a limitation that VCS depends on
the central server(s) hosting the repository.

Departing from this centralized approach, the blockchain technology premises
that we can share an immutable history of transactions without a central
server. Based on the blockchain technology, we
propose a new, collaborative 3D modeling system.
Our approach has the following features:
\begin{itemize}
    \item Decentralized, immutable, and Byzantine fault tolerant, because our approach is based on
	    the blockchain.
    \item Modular, since the modeling part and the blockchain part of our
	    approach are separated, so we can easily use various existing
		modeling tools.  
\end{itemize}

We think that our approach is useful for server-less collaborative modeling
projects, especially community based modeling projects. We expect our research
to contribute to the modeling industry, and show that the blockchain can be
used for various purposes even in graphics and modeling fields.

\begin{figure}[t]
    \begin{center}
        \includegraphics[width=0.45\textwidth]{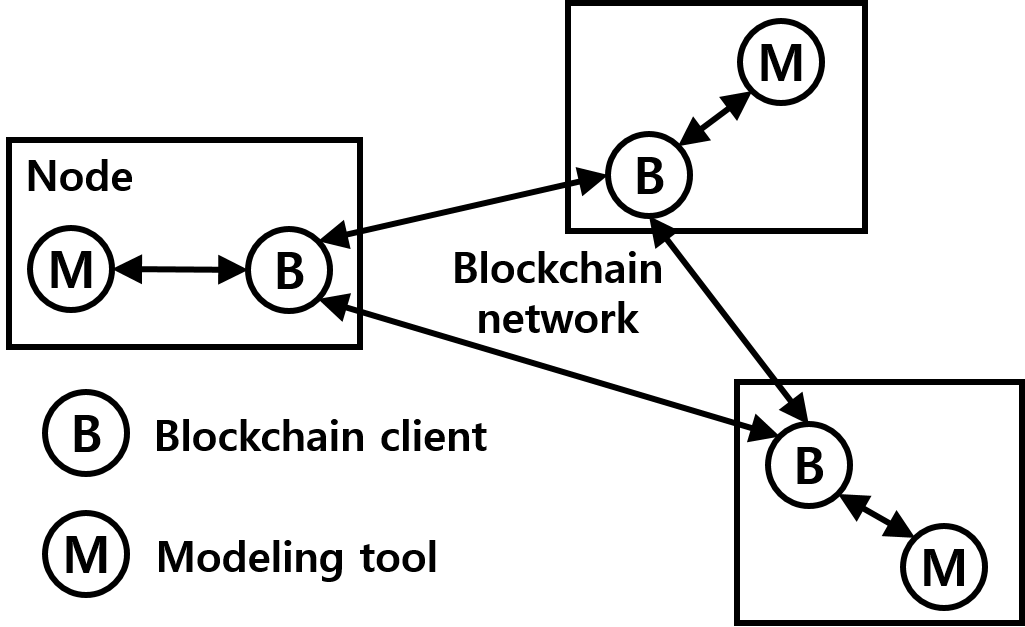}
        \caption{This figure shows a blockchain network of our approach consisting of blockchain clients, each of which is associated with a modeling tool.}
        \label{fig:Network}
    \end{center}
\end{figure}

\section{Related work}

\myParagraph{Collaborative modeling.} Dobo\v{s} and
Steed~\cite{Dobos:2012:DIA:2407746.2407766} presented an interactive
visualization tool for mesh difference and conflict resolution. Also, a
practical algorithm for diffing and merging polygonal meshes was
presented~\cite{Denning:2013:MDM:2461912.2461942}. Salvati et
al.~\cite{Salvati:2015:MCM:2816795.2818110} proposed a real-time collaborative
editing system of low-polygonal and subdivision meshes.\\

\myParagraph{Blockchain-based revision control.} There was a Kickstarter
project \cite{Gitchain}, which tries to combine concepts of Git and blockchain.
It attaches blockchain as a local proxy to Git for using Git without
depending on Git hosting servers.\\

\myParagraph{Blockchain-based database.} A recent project attaches blockchain
on a set of nodes where each node has its own MongoDB database to get the
advantages of database and blockchain together~\cite{BigchainDB}.\\

While previous studies \cite{Denning:2013:MDM:2461912.2461942}
\cite{Dobos:2012:DIA:2407746.2407766} have focused on versioning meshes for
collaboration modeling, 
our approach focuses on providing a distributed network that combines modeling
tools with a decentralized database, allowing those prior studies to be used
for server-less collaboration.

\section{Collaborative Modeling using Blockchain}

In this section, we first talk about our blockchain network, followed by various operations that our approach support.

\subsection{Network}

Our idea for collaborative 3d modeling is
to regard the modification of the mesh as a transaction
and use the blockchain as a database of modification history.
In the network, each node consists of one or modeling tools,
e.g., Blender and Maya, and a blockchain client (Fig. \ref{fig:Network}).
The blockchain client is responsible for
distributing the transactions to the network,
managing the blockchain,
and delivering the modifications from other nodes to the modeling tool.

One may consider to tightly integrate the modeling tool to the blockchain or
vice versa.  Instead of this integration approach, we choose the current
approach of separately having modeling tool and blockchain client, since this
decoupled approach enables users to use any existing modeling tool as a plugin
or to use multiple modeling tools together.
\subsection{Operations}

Our approach provides three operations: Commit, Mine, and Checkout.\\

\myParagraph{Commit a mesh.} When a user edits a mesh and requests \myEmph{commit},
the modeling tool sends the mesh to the blockchain client.
The blockchain client packs the mesh into a transaction.
Instead of storing the whole mesh data, we store the difference between the
current mesh and the prior mesh, which corresponds to the last transaction
stored in
the chain.
We can retrieve the mesh later
by iterating through the path from the root to the current transaction. This
approach makes the transaction size a lot smaller than storing the whole mesh,
so we can 
enable efficient collaborative modeling.

\begin{figure}[t]
    \begin{center}
        \includegraphics[width=0.45\textwidth]{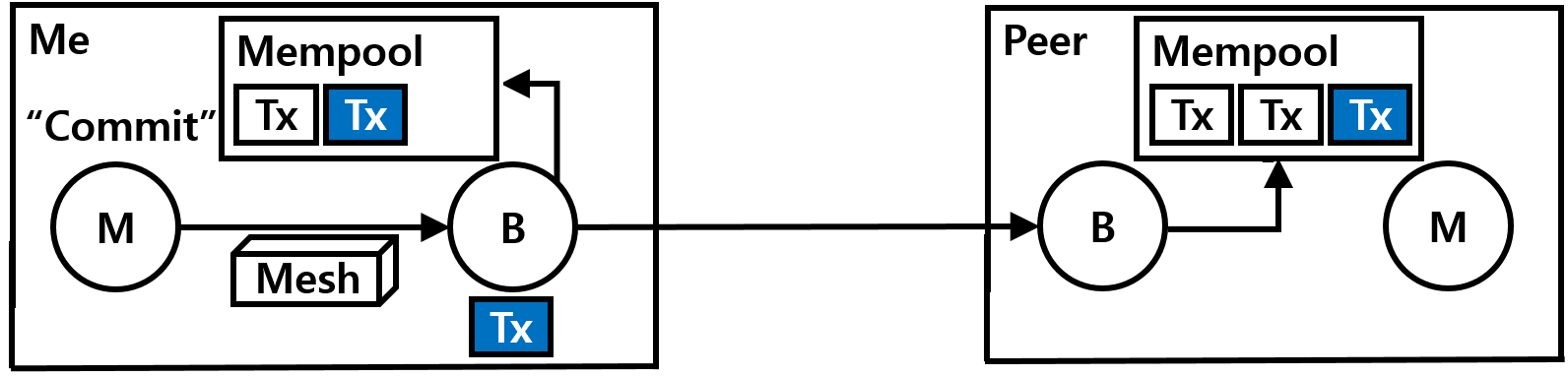}
	    \caption{Process of \myEmph{commit}. The blockchain client creates
	    the new transaction (shown in blue) using the committed mesh and
	    broadcasts it. Each blockchein client appends the transaction to
	    its mempool.
	    }
	    \label{fig:OperationCommit}
    \end{center}
\end{figure}

The blockchain client broadcasts the transaction to
the peers, and each peer stores the transaction
in  a temporary storage, \myEmph{mempool}, for the transactions waiting for
being included in a block
(Fig. \ref{fig:OperationCommit}).

\begin{figure}[b]
    \begin{center}
        \includegraphics[width=0.45\textwidth]{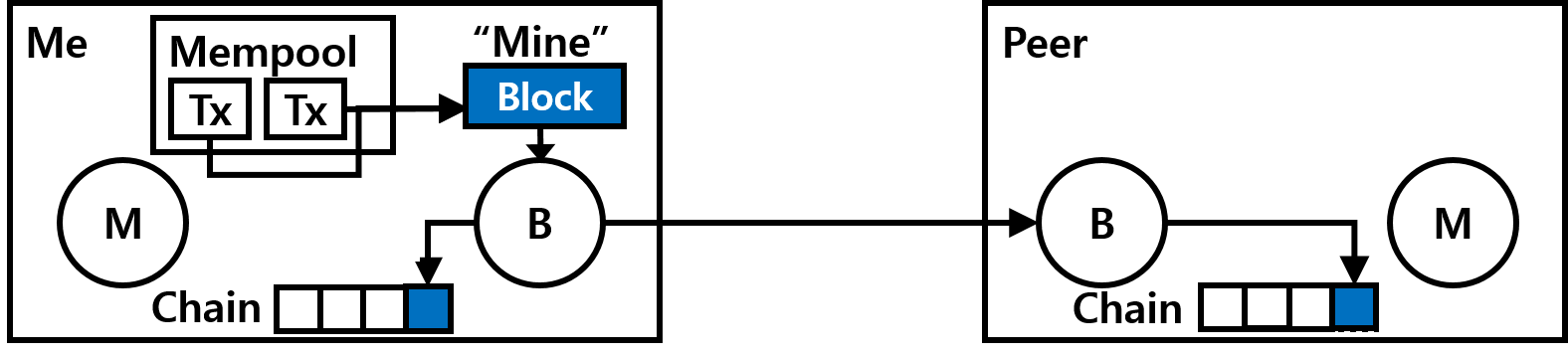}
	    \caption{Process of \myEmph{mine}. The newly created block is shown in blue.
	    }
        \label{fig:OperationMine}
    \end{center}
\end{figure}

\myParagraph{Mine a block.}
When a user requests \myEmph{mine}, the blockchain client packs the
transactions in the mempool and mines a new block.  It updates its chain and
broadcasts the block to the peers.  When the peers receive the block, they
update their own chain (Fig. \ref{fig:OperationMine}).

\myParagraph{Checkout a mesh.} When a user selects a block from the chain, and selects a transaction from the block and requests \myEmph{checkout},
the blockchain client constructs the mesh corresponding to the transaction
and send the mesh to the modeling tool.
This operation provides user with a version control system.
The user can download the latest commit,
or rollback to one of the previous commits.

\section{Implementation and Results}

\begin{figure}[t]
    \begin{center}
        \includegraphics[width=0.5\textwidth]{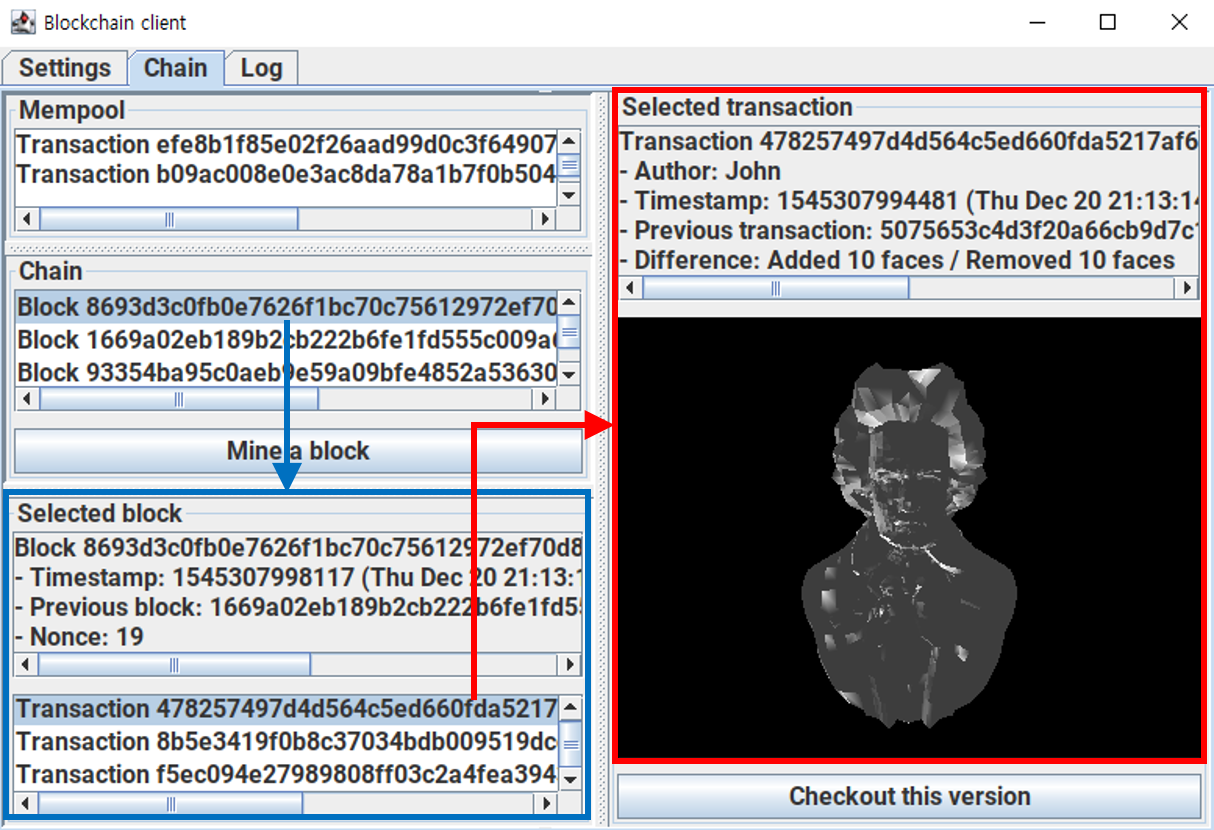}
        \caption{Screenshot of the blockchain client. User can select a block from the chain (shown in blue) and select a transaction from the block (shown in red).}
        \label{fig:Implementation}
    \end{center}
\end{figure}

We have implemented our approach by using Blender and a simple
PoW (Proof-of-Work) based blockchain written in Kotlin (Fig.
\ref{fig:Implementation}).
We used HTTP for implementing communication between Blender and the blockchain client, and communication between the blockchain clients.
The blockchain client provides a simple GUI, which shows the mempool and the chain.
The user can select a block from the chain and select a transaction from the block.
When the user selects a transaction, the blockchain client provides 3D preview
of mesh, which corresponds to the transaction.
Our implementation can be run on multiple localhost nodes in a single computer, or multiple computers.

You can see the video of how our implementation works at 
\url{https://youtu.be/s33wzSrMXwI}. You can see the code of our implementation at \url{https://github.com/Avantgarde95/C3DMB}.

\bibliographystyle{plain}
\bibliography{Refs}
\end{document}